\documentstyle[12pt]{article}
\begin{document}

\medskip

\begin{center}
{\large {\bf Nucleon number dependence of the onset of anomalous $J/\psi$
suppression and the dynamics of nuclear collisions}}
\end{center}

\medskip

\begin{center}
 Anna Nogov\'a${}^{a)}$, Neva Pi\v{s}\'utov\'a${}^{b)}$, and J\'an Pi\v{s%
}\'ut${}^{b)}$ 
\end{center}

\medskip

\begin{center}
${}^{a)}$ Institute of Physics of the Slovak Academy of Science,

SK-84228 Bratislava,Slovakia

${}^{b)}$ Department of Physics, Comenius University,

SK-84248 Bratislava, Slovakia
\end{center}

\medskip

{\bf Abstract} We point out that the data on the onset of anomalous $J/\psi$
suppression as a function of nucleon numbers
 A and B could provide information on the
dynamics of nuclear interactions. In particular the models of  anomalous $%
J/\psi$ suppression by Blaizot and Ollitrault (BO) and by Kharzeev,
Lourenco, Nardi and Satz (KLNS) are based on different assumptions on the
dynamics of nuclear collisions and lead to different predictions
of the dependence of the onset of  anomalous $J/\psi$ suppression 
on nucleon numbers of colliding nuclei. The data on this onset
as function of A and B  could become a tool for the study of the dynamics of
nuclear  collisions and bring further evidence on $J/\psi$ suppression by
new form of hadronic matter, possibly Quark-Gluon Plasma.
 In particular we propose to study $J/\psi$ suppression
in A+Pb interaction with nucleon number of A between 58 and 73 or a
 bit higher.
\section{Introduction}
\label{intro}
 Experimental data of the NA38 and NA50 Collaborations at the
CERN SPS \cite{NA50a,NA50b,NA50c,NA50d}  on $J/\psi$ suppression has shown
that the anomalous $J/\psi$ suppression is observed in near central Pb+Pb
collisions but it is absent in peripheral Pb+Pb and in interactions induced
by ions lighter (or equal) to Sulphur.

Non- anomalous suppression of $J/\psi$ has been phenomenologically well
described by the desintegration of $J/\psi$ by nucleons present in  nuclei 
\cite{GHa,GHb}.

There have appeared numerous attempts to describe the anomalous suppression
of $J/\psi$. We shall not try to make here a complete list of these
attempts. A review of approaches based on $J/\psi$ suppression by comovers
can be found in Refs.\cite{Vogta,ACF}. For studies based on the initial
state interactions see Refs.\cite{Hwa2,Gale1}.

A rather abrupt increase of $J/\psi$ suppression with increasing transverse
energy ($E_T$) in Pb-Pb interactions observed by the NA50 Collaboration
 \cite{NA50d} can also be interpreted
 as due to the formation of an "anomalous"
state of hadronic matter, possibly of the QGP. Models of this type have been
suggested by Blaizot and Ollitrault (referred to below as BO) \cite{BO} and
by Kharzeev, Louren\c{c}o, Nardi and Satz (KLNS) \cite{KLNS}. Both models have
built in the onset of the formation of matter in the "anomalous" state so
they are both able to describe the onset of the anomalous $J/\psi$
suppression at a given value of $E_T$ in Pb-Pb interactions at the CERN SPS.

The purpose of this note is to point out that BO and KLNS models are based
on  different assumptions on dynamics of 
 production of matter with high  energy
density in nuclear collisions and if they were the only possibilities it
would be possible to learn which one gives
 the correct description by studying
the occurence of anomalous $J/\psi $ suppression as function of nucleon
numbers of colliding nuclei. Since, in fact, BO and KLNS are not the only
possibilities, the data on the dependence of
 anomalous $J/\psi $ suppression 
on nucleon numbers of colliding nuclei, if available, would
provide an information on the mechanism of transverse energy production in
nuclear collisions. In the Sect.2 we shall discuss this point in more
detail, in Sect.3 we shall 
analyze the possibility of discriminating between the BO and KLNS models by
using suitable experimental data and in  Sect.4
we shall present  conclusions and comments.

\section{Nucleon number dependence of the onset of anomalous $J/\protect\psi$
production in BO and KLNS models}

\label{density} In the present paper we shall use a simple model of nuclei
as spheres with constant nucleon densities and radii of
 $r_A=1,2A^{1/3}$fm 
where A is the nucleon number. In this simplified description some results
can be  obtained in a transparent analytical form.

Both BO \cite{BO} and KLNS \cite{KLNS} assume that the mechanism  of
anomalous $J/\psi$ suppression observed by the NA50 Collaboration  at the
CERN SPS is due to the dissolution of $J/\psi$ by
the Quark-  Gluon Plasma
(QGP). The time spent by the $J/\psi$ in the QGP does not enter into these
models what means that the dissolution is assumed to be immediate.
The dynamics of at least the first decisive  stage of the nuclear collision
is assumed to be longitudinal  in the sense that the criterion for the
formation of QGP depends  only on the particular tube- on- tube
interaction. Following the picture of the Wounded Nucleon Model \cite{WNM}
BO introduce  density of participating nucleons per unit transverse area 
\begin{equation}
n_p(\vec s,\vec b)=T_A(\vec s)[1-exp(-\sigma_NT_B(\vec s-\vec b))] +
 T_B(\vec s-\vec b) [1-exp(-\sigma_NT_A(\vec s))]
\label{eq.1}
\end{equation}
where $\sigma_N$ is the non-diffractive nucleon- nucleon
 cross- section, $\vec b$
is the impact parameter and $\vec s$ gives the transverse
 position of a nucleon
with respect to the center of the nucleus $A$.
 Functions $T_A(\vec s)$
  and 
$T_B(\vec s-\vec b)$ give the nucleon densities per
 unit area in the transverse plane
with 
\begin{equation}
T_A(s)=\int_{-\infty}^{+\infty}\rho_A(s,z)dz
\label{eq.2}
\end{equation}
The condition for the formation of QGP at a particular value
 of $\vec b$ and $\vec s$
is then stated in Ref.\cite{BO} as 
\begin{equation}
n_p(\vec s,\vec b)\ge n_{BO,c}
\label{eq.3}
\end{equation}
where $n_{BO,c}$ is supposed to be just equal  to the maximal possible
value reached in $S+U$ collisions.
Note that the condition in Eq.(3) makes no explicit mention of the role
possibly played by the formation times of the secondary hadrons or
partons, which may
be of some importance, see e.g. Ref.\cite{PPZ}.

Neglecting exponential factors in Eq.(1), using the sharp sphere
approximation and multiplying $n_p(\vec s,\vec b)$ by $\sigma_N=$30mb
 to get a dimensionless quantity
 we obtain  Eq.(3) in the following form
\begin{equation}
\kappa_{BO}\equiv \rho_0\sigma_N2\sqrt{R_A^2-s^2}+
 \rho_0\sigma_N2
 \sqrt{R_B^2-(\vec s-\vec b)^2} \ge \kappa_{BO,c}
 \label{eq.4}
\end{equation}         
where $\kappa_{BO}$ is defined as $n_p$ multipied by $\sigma_N$ and
$\rho_0=0.138fm^{-3}$.
Following their earlier work \cite{BOe} BO assume
 in Ref.\cite{BO} that all 
$J/\psi$'s produced at values of $\vec b,\vec s$
 satisfying Eq.(4) are completely suppressed
(independently of whether they are produced directly or via e.g. radiative
decays of $\chi$). When discussing the onset of QGP formation and thus also
the onset of  anomalous suppression of $J/\psi$ as function of nucleon
numbers  $A$
and $B$  we have to consider the collision
of tubes  with $b=0$ and $s=0$, obtaining
thus 
\begin{equation}
2\rho_0\sigma_N 1.2(A^{1/3}+B^{1/3})\ge \kappa_{BO,c}
  \label{eq.5}
\end{equation}
which in variables $x=A^{1/3}$, $y=B^{1/3}$ gives
\begin{equation}
x+y=1.006\kappa_{BO,c}
\label{eq.6}
\end{equation}
In their model of the formation of QGP KLNS \cite{KLNS} use also the 
picture of tube- on- tube collisions, but in contradistinction to  BO \cite
{BO} they use as a criterion the interaction density \cite{K} defined (in
the sharp sphere model) as 
\begin{equation}
\kappa_{K}= \frac{N_c}{N_w}= \frac {\rho_0\sigma_N2\sqrt{R_A^2-s^2}.
\rho_0\sigma_N2\sqrt{R_B^2-(\vec s-\vec b)^2}}
 {\rho_0\sigma_N2\sqrt{R_A^2-s^2}+
\rho_0\sigma_N2\sqrt{R_B^2-(\vec s-\vec b)^2}}
\label{eq.7}
\end{equation}
The onset of $J/\psi$ suppression at the highest available $E_T$ (neglecting
fluctuations in $E_T$) corresponds again to $b=0$, $s=0$ and Eq.(7) in this
case leads to
$$
\frac{xy}{x+y}=1.006\kappa_{K,c}
$$
what is equivalent to
\begin{equation}
y=\frac{a^2}{x-a}+a, \qquad a=1.006\kappa_{K,c}
\label{eq.8}
\end{equation}
 As pointed out by KLNS  the $\kappa_{K}$ as defined by the first part of
Eq.(7) can in principle be determined experimentally. The number of nucleon-
nucleon collision $N_c$ is proportional to the Drell- Yan (DY) pair
production, whereas the number of secondary hadrons produced is proportional
to the number of participating (wounded) nucleons. At the CERN SPS this
statement is only approximately correct, since the data on transverse energy
($E_T$) distributions and in particular the data on $E_T$ distributions
associated to DY production do require some contribution to $E_T$
proportional to $N_c$, for details see the paper by Armesto, Capella, and
Ferreiro \cite{ACF} based on the Dual Parton Model \cite{DPM}.

We would like to stress that the dynamics responsible for energy density as
given by Eq.(7) is rather different from that corresponding to Eq.(4). When
considering the A+B interaction in the cms of individual nucleon- nucleon
collisions, the numerator in Eq.(7)
 is proportional to the number of nucleon- nucleon
collisions (within a given tube-on-tube collision)
 whereas the denominator is proportional to the sum of the
lengths (and volumes)
 of the two interacting tubes. So one of possible interpretations of
Eq.(7) is that $E_T$ relevant for the QGP formation is due to semihard 
gluon- gluon interactions, since the repetition of soft ones would be 
damped by the Landau- Pomeranchuk effect.
 The ratio in Eq.(7) is then proportional to the total
 transverse energy over the interaction volume and thus to energy density.

For the sake of completeness we shall now present a few details concerning
 the calculations. When using the 
 Gerschel and  H\"ufner \cite{GHa,GHb} mechanism we have taken
 $\sigma_{abs}=$7mb and we have not introduced
the effects  due to the fluctuations of the $E_T=E_T(b)$ dependence. 
The calculations has been performed by using the expression 
\begin{equation}
S=\frac {N}{N_0}
\label{eq.9}
\end{equation}
where 
\[
N_0=\int^{R_A}_0sds\int^{2\pi}_0d\theta
 2\sqrt{R_A^2-s^2}2L_B(\vec b,\vec s,\theta)
\]
\[
L_A(s)=\sqrt{R^2_A-s^2},\quad L_B(\vec b,\vec s,\theta)=
\sqrt{R_B^2-b^2-s^2+2bscos(%
\theta)} 
\]
when the expression under the square-root in $L_B$ is positive, and $%
L_B(\vec b,\vec s,\theta)=0$ when this expression is negative.

The numerator is given as 
\[
N=\int_0^{R_A}sds\int_0^{2\pi}d\theta \Theta(R_B^2-b^2-s^2+2bscos(\theta)) 
\]
\[
\int_{-L_A(s)}^{L_A(s)}dz_A\int_{-L_B(s,\theta)}^{L_B(s,\theta)}dz_B
e^{-\rho_A\sigma_a[z_A+L_A(s)]} 
\]
\[
e^{-\rho_B\sigma_a[z_B+L_B(s,\theta)]}\Theta(\kappa_c-\kappa) 
\]
where we have put $\sigma_a= 7$mb. The last term is to be interpreted as $%
\Theta(\kappa_{c,K}^{\chi}-\kappa_{K}^{\chi})$ or $\Theta(\kappa_{c,BO}-%
\kappa_{BO})$ respectively, or in the case of the two threshold scheme
we have to make
the replacement, see Ref. \cite{KLNS} 
\[
\Theta(\kappa_c-\kappa)\rightarrow 
0.4\Theta(\kappa_{K,c}^{\chi}-\kappa_{K})
+0.6\Theta(\kappa_{K,c}^{\psi}-\kappa_{K}) 
\]
where the first threshold corresponds to the dissolution  of $\chi$
by QGP and the second one to the  dissolution of $J/\psi$.

\section{On the possibility of discriminating between BO and KLNS models by
data}

\label{dis}
In this section we shall show first that KLNS and BO schemes cannot be
made equivalent in what concerns the
dependence of anomalous $J/\psi$ suppression on $E_T$ and on nucleon
numbers A,B of colliding nuclei. As an illustration we shall consider
a situation in which the onset of anomalous $J/\psi$ suppression
is assumed to occur (without taking into account the fluctuations
of $E_T$) just in the interaction of the two longest tubes in S+U
collision. In this academic example we take U-nucleus as spherical.
In this case $\kappa_{BO,c}=9.312$ and $\kappa_{K,c}=2.086$ and
the curves for the onset of the anomalous $J/\psi$ suppression in the
$x=A^{1/3}$ and $y=B^{1/3}$ plane are given by Eqs.(6,8) and this leads to
\begin{equation}
(BO):\quad y=9.37-x, \qquad (KLNS):\quad y=\frac{4.41}{x-2.1}+2.1
\label{eq.10}
\end{equation}
The two curves are shown in Fig.1. As expected the two curves cross each
other for values of $x$ and $y$ corresponding to S and U nuclei.

With $\kappa_{BO,c}$=9.312 and $\kappa_{K,c}$=2.08 fixed by the assumption
that anomalous $J/\psi$ suppression sets on just at the end of the $E_T$
distribution for S-U interactions we can now study the onset of anomalous
$J/\psi$ suppression in Pb-Pb collisions. In Fig.2 we present $J/\psi$
suppression calculated via Eq.(9) in Pb-Pb interactions. We have calculated
$E_T(b)$ by using the relationship $E_T(b)=$0.325$N_w(b)$GeV, where
 $N_w(b)$ is the number of wounded
 (participating) nucleons at a given value of $b$.

Fig.2 demonstrates an interesting, although very simple effect. When
$\kappa_{BO,c}$ and $\kappa_{K,c}$ are fixed by the condition related
to the interaction of the two longest tubes in S-U interaction, the
onset of the anomalous $J/\psi$ suppression in Pb-Pb collisions appear
at different values of $E_T$ in BO and KLNS schemes.

Since, in fact, the anomalous $J/\psi$ suppression has not been observed
in S-U interactions we find it more appropriate to determine the values
$\kappa_{BO,c}$ and $\kappa_{K,c}$ by the data \cite{NA50d} on $J/\psi$
suppression in Pb-Pb collisions.

In Fig.3 we present the NA50 Collaboration data \cite{NA50d} 
on  the
survival  probability of $J/\psi$ as a function of the transverse energy
 $E_T$  in Pb-Pb interactions at 158 GeV per nucleon and three curves 
giving $J/\psi$ suppression under different conditions: the curve denoted
as "no anom."
corresponds to pure Gerschel- H\"{u}fner mechanism with $\sigma_{abs}$=
7mb, the curve "BO" is given by the BO scheme with $\kappa_{BO,c}$=9,75 and
with all $J/\psi$'s with $\kappa_{BO}\ge \kappa_{BO,c}$ completely
absorbed, the curve "KLNS" corresponds
 to the KLNS scheme with $\kappa_{K,c}$=2,43
 and with all $J/\psi$'s with $\kappa_K \ge \kappa_{K,c}$ completely
absorbed.

 In their work KLNS \cite{KLNS} take into account that about 40\% of
$J/\psi$'s observed in nuclear collisions is due to radiative $\chi$-
decays and assume that $\chi$ is suppressed at a different threshold
in $\kappa_K$ than  directly produced $J/\psi$. This permits them to
describe also the second decrease of survival probability of $J/\psi$
in the region of $E_T>$100GeV indicated by the data.

The same two threshold mechanism could also be implemented into the BO
scheme and in this way both BO and KLNS schemes could describe the
Pb+Pb data in a similar way. Within our simple model this could be
achieved by increasing the value of $\sigma_{abs}$ and lowering the
tresholds given by $\kappa_{K,c}$ and $\kappa_{BO,c}$. We shall not follow
this path here since the purpose of this note is to study only the onset
(the first one) of the anomalous $J/\psi$ suppression and not the detailed
shape of the survival probabability, including the possibility of a second
threshold (the possible onset of $\chi$ suppression).

 As can be seen in Fig.3 shapes of $S(E_T)$
 obtained within BO and KLNS schemes in Pb-Pb interactions
 are rather similar when $\kappa_{BO,c}$ and $\kappa_{K,c}$ are fixed
 by the condition that the onset of anomalous suppression starts at
 the same value of $E_T$. It seems therefore that  by using only the
 Pb-Pb data it would be very difficult to discriminate between the
 two schemes.

In the rest of this section we shall try to find out whether one can 
discriminate between the BO and KLNS by studying the onset of anomalous $%
J/\psi$  suppression as the function of atomic numbers A and B of the
colliding  nuclei. Note that we are neglecting the role of possible
fluctuations  in the energy density of matter formed in nuclear collisions, 
see Ref.\cite{BO2}.

In Fig.4 we show curves giving the onset of anomalous $J/\psi$ suppression
in the $x=A^{1/3}$ and $y=B^{1/3}$ plane corresponding to values of
$\kappa_{K,c}=2.43$ and $\kappa_{BO,c}=9.75$ (the values used also in
calculations leading to Fig.3).

As can be seen in Fig.4 for A+Pb collisions with 57${}\le A\le{}$73
 the survival
probability as function of $E_T$ shows the presence of anomalous
suppression within the BO scheme and no anomaly for the KLNS one.
For nuclei with $74\le A$ both KLNS and BO show the presence of anomaly
but the anomaly sets on at different values of $E_T$. As an illustration
we show in Fig.5 the results obtained again via Eq.(9) with
$\kappa_{BO,c}=$9.75 and $\kappa_{K,c}=$2.43 for the $As^{75}$-Pb
interactions, still using $E_T(b)=$0.325GeV$N_w(b)$.

\section{Comments and conclusions}

As can be seen in Fig.4 there do exist regions
 in the A,B plane where the BO
picture of $J/\psi$ suppression based on the Wounded Nucleon Model \cite{WNM}
predicts anomalous $J/\psi$ suppression whereas the KLNS scheme
 \cite{KLNS} predicts its absence.
 For Pb as one of colliding nuclei this region
for the other nucleus A contains nuclei with 
\begin{equation}
57\le A \le 73  \label{eq.11}
\end{equation}
including nuclei as $Ni_{59}$, $Cu_{64}$ and $Zn_{65}$.
For A${}\ge 74$ the anomalous suppression is present in both schemes
 but sets on
 at lower value of $E_T$ in the BO case.

To conclude: we propose to study experimentally anomalous $J/\psi$
suppression at the CERN SPS in collisions of A-Pb or Pb-A with the
nucleon number A within the interval given by Eq.(11).
 The presence of anomalous $J/\psi$
suppression for A-Pb collisions with A within the interval given by Eq.(11)
 would support the scheme of high energy density formation
suggested by Blaizot and Ollitrault \cite{BO}, the absence of it would give
evidence in favour of the Kharzeev, Louren\c{c}o, Nardi and Satz
 model \cite{KLNS}.

A study of the onset of anomalous $J/\psi$ suppression in a series of A-Pb
interactions including the available data on S-Pb and Pb-Pb,
and some of A-Pb with A given by Eq.(11) (or with A only a bit higher)
 at 158Gev per nucleon
 could bring interesting
evidence on the dynamics of production of critical energy density 
for QGP formation in this energy range.
 Moreover, since at present
the anomalous $J/\psi$ suppression has been observed in only one case,
the second observation in A-Pb with a lighter A would support 
the idea of anomalous $J/\psi$ suppression by an anomalous state of
matter, possibly by QGP. As seen in Fig.5 the difference between
predictions of BO and KLNS schemes is not very  large and some indeterminacy
will be introduced by $E_T$ fluctuations \cite{KLNS,BO2,CFK} not
considered here we still believe that a study of A-Pb interactions
with A given by Eq.(11) or nearby could bring in a very useful
information.

{\bf Acknowledgements}: the authors are indebted to A.Capella and R.Hwa for
useful discussions on the problem of dynamics of nuclear collisions and of
anomalous $J/\psi $ suppression.The present work has been supported in part
by the U.S.- Slovakia Science and Technology Program, under Grant 
No. 003- 95 (INT 9319091) and by the Slovak Grant Agency under Grant V2F13-G.

\bigskip

{\bf Figure Captions}

\medskip

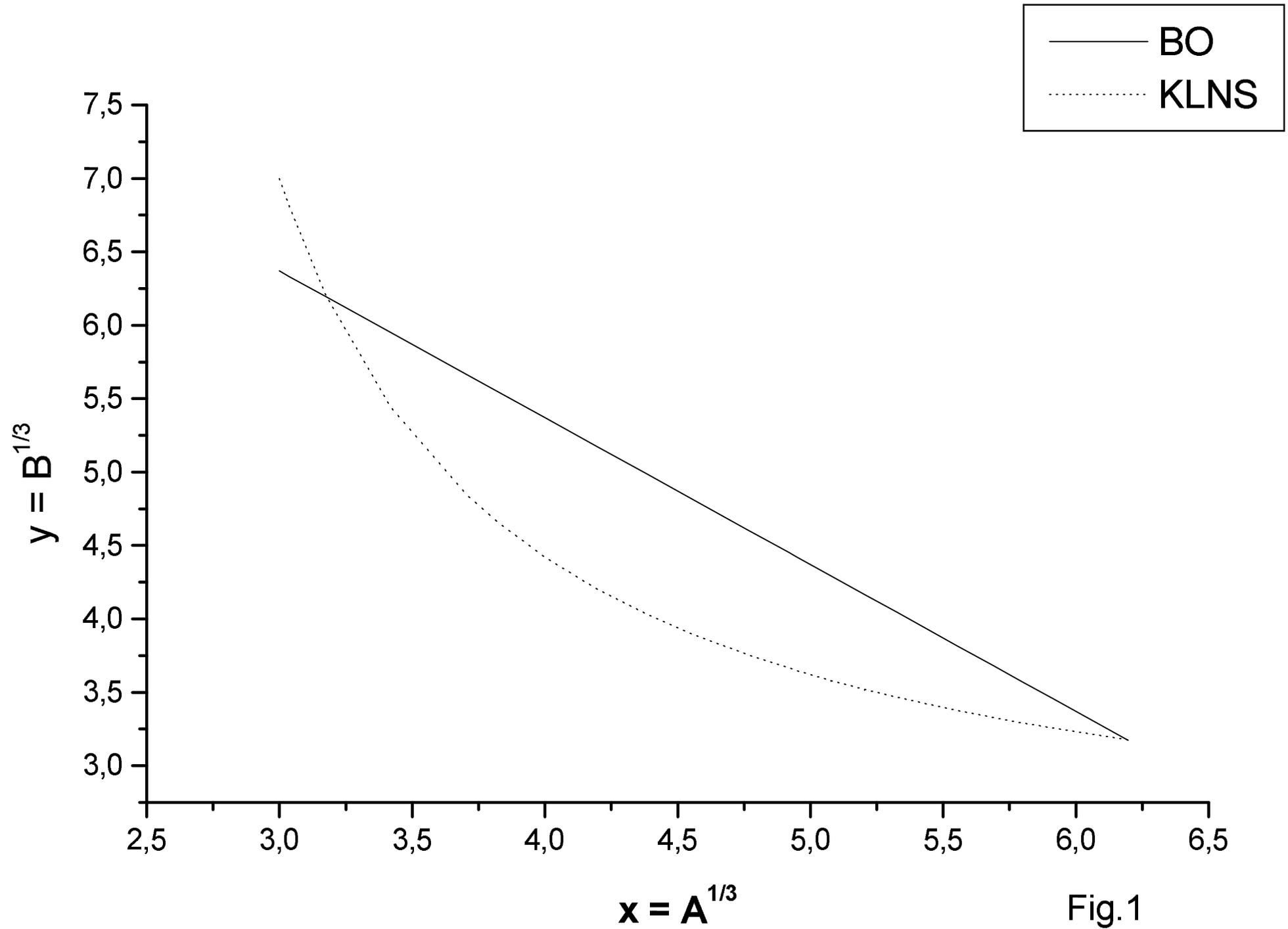
{\bf Fig.1}
Curves denoted as KLNS and BO give lines of the onset of anomalous
$J/\psi$ suppression in BO \cite{BO} and KLNS \cite{KLNS} schemes
as functions of nucleon numbers A and B. The curves are given explicitly
in Eq.(10). The critical values are $\kappa_{BO,c}$=9.312
 and $\kappa_{K,c}$=2.086.

\medskip

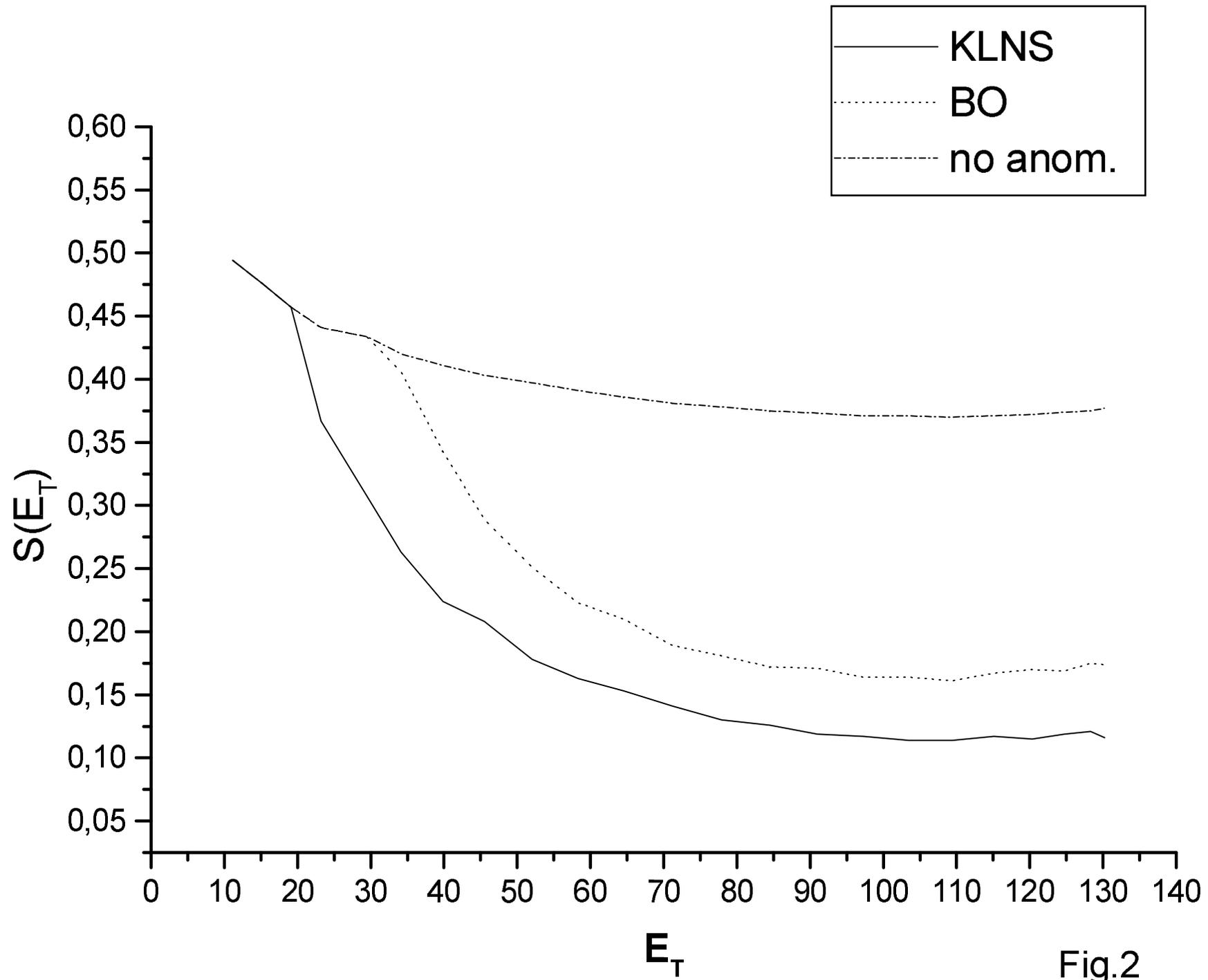
{\bf Fig.2}
Dependence of the survival probability $S(E_T)$ on transverse energy
$E_T$ calculated via Eq.(9) for KLNS and BO schemes with
$\kappa_{BO,c}$=9.312 and $\kappa_{K,c}$=2.086 (corresponding to the onset
of anomalous suppression just above the S-U interactions)
 in the space of nucleon numbers A and B. The curve denoted as "no anom."
gives the case with  no anomalous suppression.

\medskip

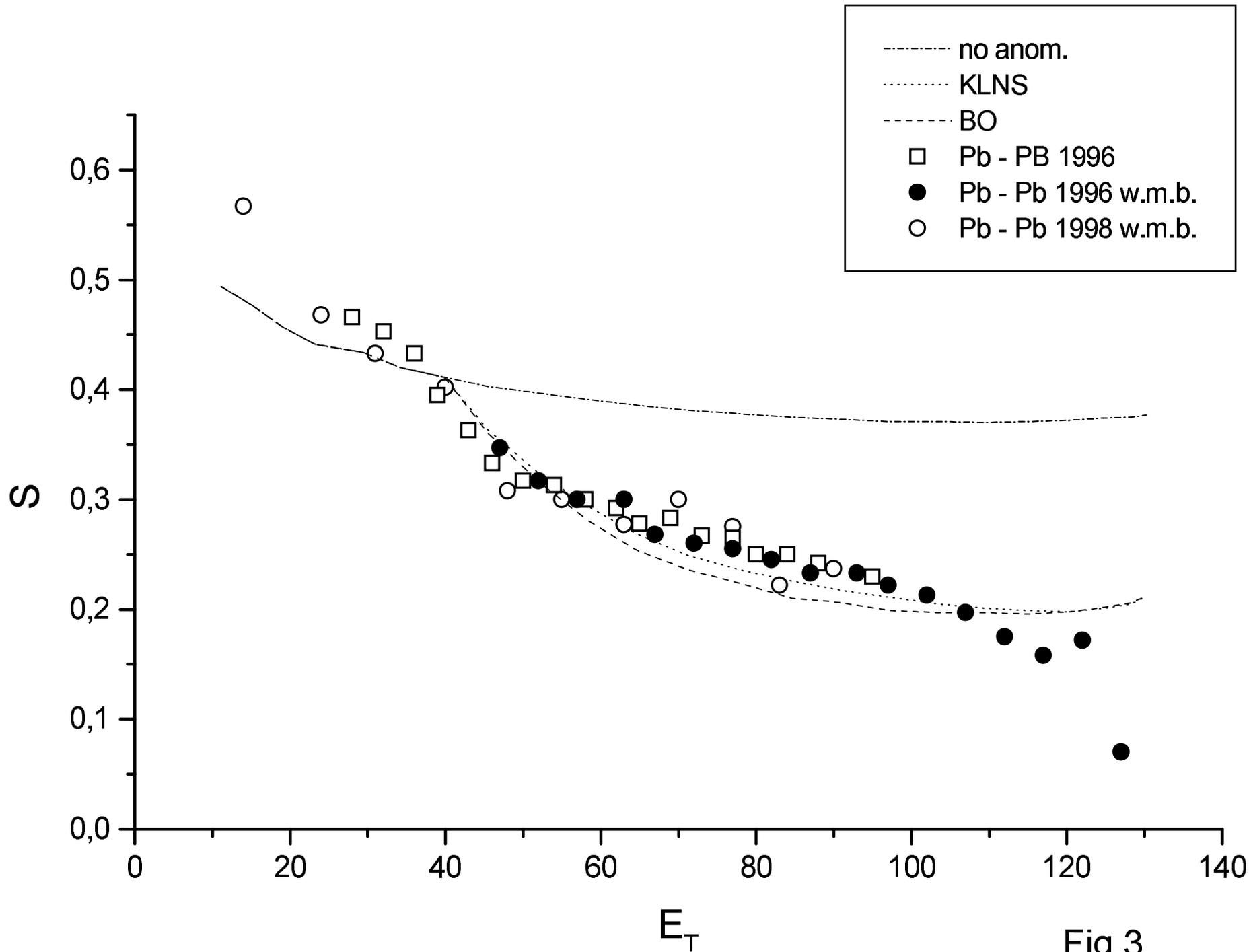
{\bf Fig.3}
Comparison of survival probabilities of $J/\psi$ in Pb-Pb collisions
 in cases of no anomalous suppression ("no anom.") and BO and KLNS
 schemes. More details are given in the text. The data are taken from
 Ref.\cite{NA50d} and w.m.b means "with minimum bias".The critical
  values of $\kappa$'s are $\kappa_{BO,c}$=9.75 and $\kappa_{K,c}$=2.43

\medskip

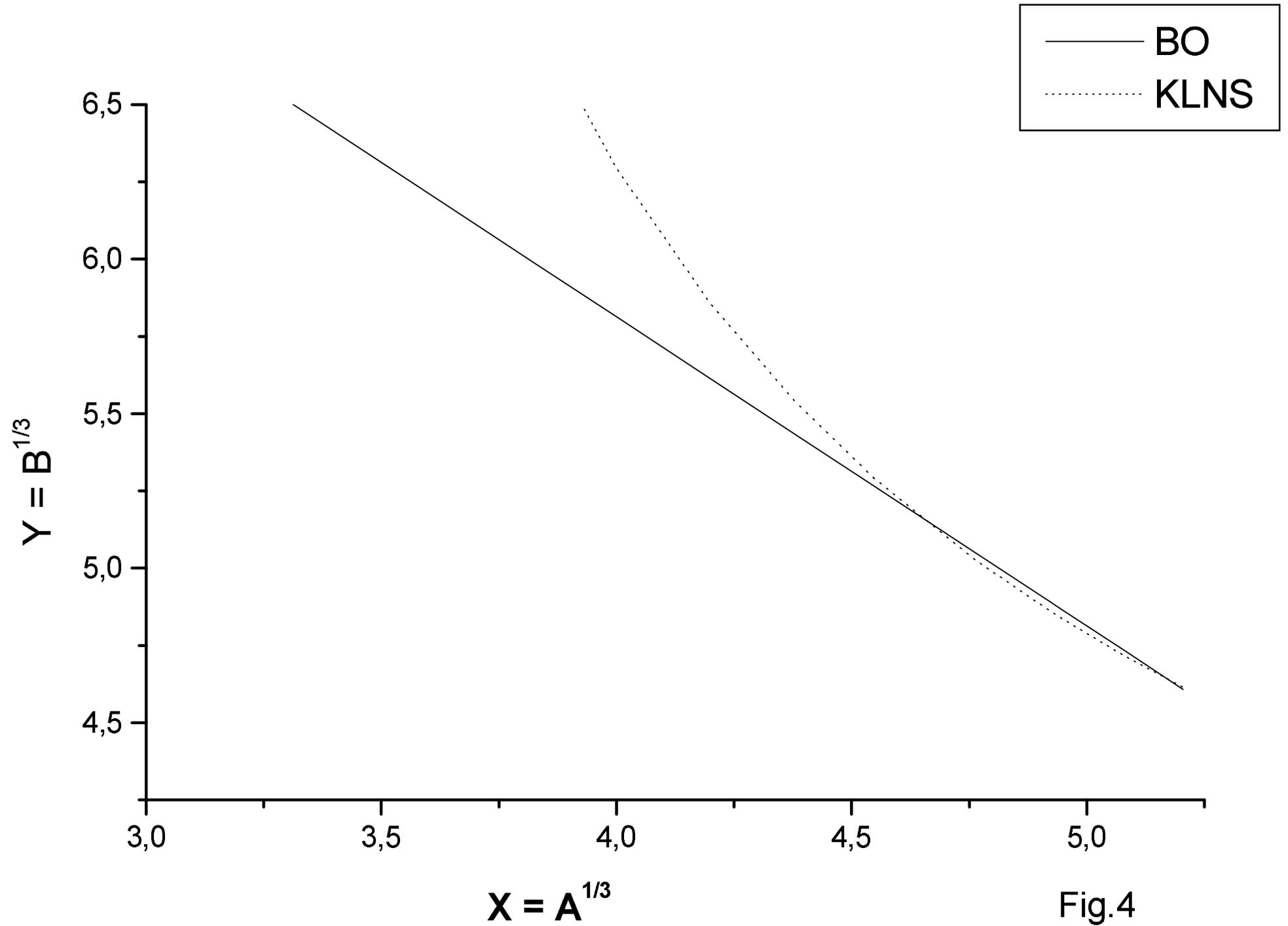
{\bf Fig.4}
Curves giving the onset of anomalous $J/\psi$ suppression within KLNS
and BO schemes with values $\kappa_{BO,c}$=9.75 and $\kappa_{K,c}$=2.43
obtained in comparison of KLNS and BO schemes with data in Fig.3.

\medskip

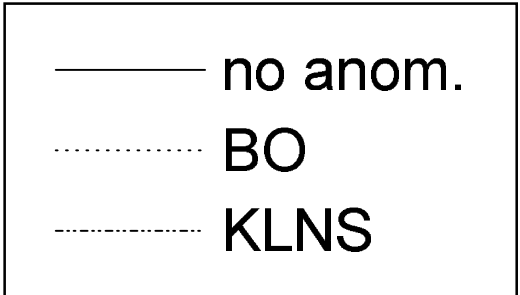
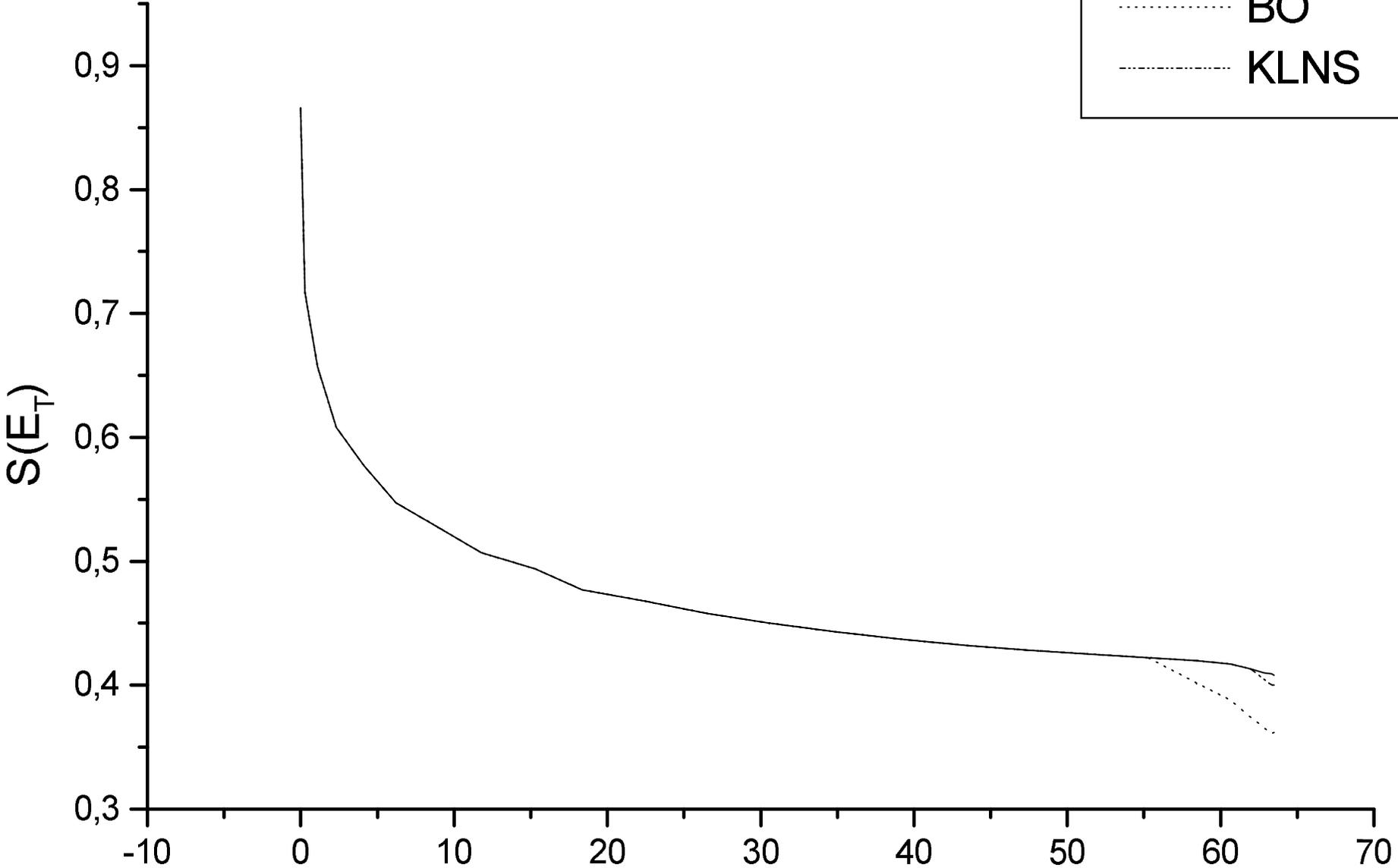
{\bf Fig.5}
Survival probability $S(E_T)$ calculated with
$\kappa_{BO,c}$=9.75 and $\kappa_{K,c}$=2.43 for As${}^{75}$-Pb interactions
for BO and KLNS schemes.

\end{document}